# >5% Compressive Strain in Graphene via the Self-Rolled-up Membrane Platform

Paul Froeter,[1] Parsian K. Mohseni,[1,2] Apratim Khandelwal,[1] and Xiuling Li[1,2*]

[1]Department of Electrical and Computer Engineering, University of Illinois, Urbana, IL 61801

[2]Current address: Department of Electrical and Computer Engineering, Rochester Institute of Technology, Rochester, NY 14623

[3]Department of Electrical and Computer Engineering, University of Texas, Austin, TX 78738

*Corresponding author: Xiuling.li@utexas.edu

**Abstract**

Graphene is an atomically thin metallic membrane capable of sustaining reversible strain and offers a tempting prospect of controlling its optoelectronic properties via strain. Graphene's exceptional mechanical flexibility (Young's modulus) and tensile strength provide a lot of room for strain engineering. Here we use the self-rolled-up membrane (S-RuM) platform for strain engineering and integration of graphene with stressed dielectric (e.g. $SiN_x$) thin films. Graphene rolls up or down together with the stressed film upon releasing from the substrate and the curvature of the rolled-up film stack enables the strain tuning of the graphene monolayer. Raman spectroscopy was used to characterize the uniaxial strain in rolled-up graphene by quantifying the red-shift and splitting of the G peak ($G^+$ and $G^-$) in the doubly degenerate $E_{2g}$ optical mode. ~5% compressive strain is realized using a S-RuM diameter of ~ 2 μm. By reducing the diameter of the S-RuM structure, even higher strain level can be reached. The S-RuM approach can also be readily applied to induce strain in other materials beyond the level that can be achieved using conventional approaches.

**Introduction**

Graphene is a 2D sheet of $sp^2$ hybridized carbon atoms arranged in a honeycomb lattice. Since its discovery it has attracted a lot of attention due to its unique mechanical, electrical and optical properties [1–3]. Moreover, the atomically thin nature of graphene provides several knobs to tune its properties based on the required application. One such unique knob is the use of long-range strains to tune its opto-electronic properties like carrier density, band gap and work function [3–6]. Modulation of such effects has been studied in great detail by both first-principles calculations and experiments [7–9]. While most experimental techniques have introduced platforms that can study either one or two such effects individually, here-in we show that our platform can be used to study multiple effects simultaneously. Due to lattice mismatch [10] or the substrate roughness [11,12], strains are expected to arise naturally in graphene. However, there are several ways to intentionally induce and control uniaxial [7,13,14] and biaxial [9,15] strain in graphene. Most of the techniques involve complex fabrication procedures and thus suffer from degraded graphene quality hindering the observation of strain effects. Here we report a simple strain tuning method by using $SiN_x$ stressed thin films as a scaffold to support and apply strain to CVD grown graphene monolayers.

In the past, a variety of material combinations including semiconductors [16,17], oxides [18], nitrides [19], metals [20] polymers [21] and other hybrid thin films [22]17,36 have been rolled-up into micro or nanotubes, scrolls, pockets and helical structures. These structures are formed by spontaneous deformation of stressed thin films driven by relaxation of strain energy. These self-rolled-up membranes (S-RuM), first discovered by Prinz et al in 2000, have proven to be an excellent platform for various on-chip and off-chip applications in electronics, optics, materials science, biology and micro/nanofluidics [23–25]. The overarching physical principle of S-RuMs is strain-driven spontaneous deformation of thin-film membranes into a cylindrical form factor. The membranes can be released from their substrates to self-assemble into tubes, coils, rings or helixes depending upon the initial 2D geometry of the patterned nanomembrane. These membranes can be rolled-up by combining standard lithography and etching processes. Most rolling mechanisms are based on the use of a sacrificial layer that can be etched via a selective dry etch [26] or wet-etch [27] process. In this study, we used a wet-etch process to achieve a controlled release of the 2D pattern (graphene/$SiN_x$) to serially tune the uniaxial strain in the CVD grown monolayer graphene. The diameter of the resulting microtubular structure is linked directly to the strain accumulated in the graphene and Raman spectroscopy is used to study the imposition of the uniaxial strain.

**Experimental**

Monolayer graphene films are grown through the Cu-catalyzed chemical vapor deposition (CVD) technique, followed by direct deposition, via the well-established poly(methyl-methacrylate) (PMMA) transfer process [28], onto strained bilayer stacks of $SiN_x$. Upon photo-lithographic definition of rectangular pads and wet-etching of an underlying sacrificial layer, an inherent strain differential arising within the bilayer stack of low- and high-frequency plasma-enhanced chemical vapor deposition (PECVD)-formed $SiN_x$ films causes the spontaneous rolling of micro-tubular structures [19]. Thus, the deposited graphene film is subjected to uniaxial compression upon release of the inherently strained $SiN_x$ membrane. The degree of strain applied to the $SiN_x$ stack and, therefore, that which is induced in the graphene sheet, is a function of the micro-tube diameter, and is directly tuned according to the relative thickness of the low- and high-frequency deposited nitride layers. Alternatively, the diameter of the micro-tubes can be controlled through partial or full etching of the underlying sacrificial layer.

**Results and Discussions**

Figure 1 shows tilted-view scanning electron microscopy (SEM) images of flat, partially released, and fully rolled graphene films on or in $SiN_x$ microtubes.

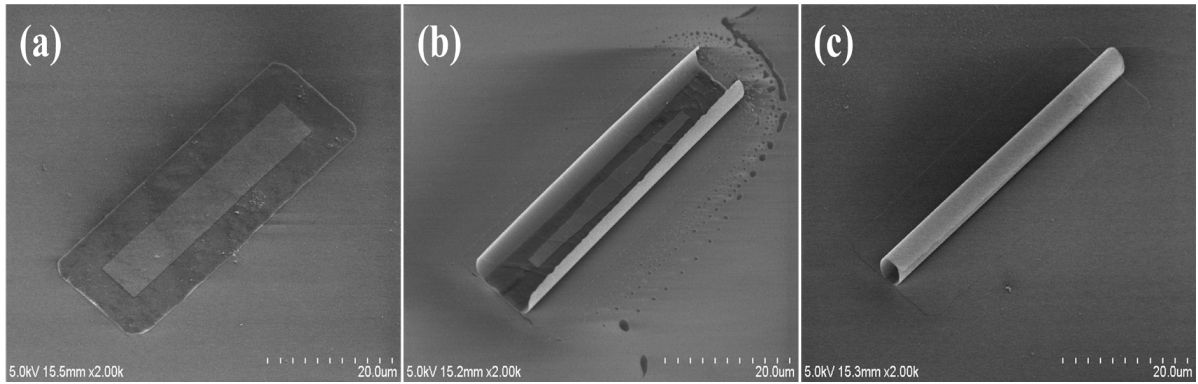

**Figure 1.** 45° titled-view SEM images of (a) un-rolled, (b) partially rolled, and (c) fully rolled graphene/$SiN_x$ micro-tube. The graphene film is situated on top of the flat (unreleased) pad, as shown in (a). However, as the microtube begins to roll-up, the graphene film becomes compressively strained and contorts in tandem with, and inside, the microtube structure, as shown in (b) and (c).

Multi-diameter chips (MDC) are very useful for this type of strain analysis as they provide a larger range of results from a single sheet of graphene, improving the accuracy of batch characterization runs. These MDC consist of a varying thickness compressive or tensile layer and multiple photolithography patterning steps, shown in Figure 2. This multi-step deposition process is guided by the equation relating bending radius to film parameters and supported in our investigation. In our experiments Mg was used as the sacrificial layer, due to its ability to act as an etch stop in CF4-based plasma, which is commonly used to define the SiNx. The desired deposition and patterning routine shown in figure 2 can be continued until all thicknesses of compressive film are met. Note that all patterns were the same dimension; therefore, larger radii micro-tubes only have ¾'s of a turn (figure 2.5j) or ½ of a turn (figure 2.5k). The LF-SiNx layers in figure 2.5i-k total in thickness of 25nm, 35nm, and 55nm, respectively, yielding 6, 6.8, and 9.5 μm. These diameters are larger than predicted most likely due to the interaction of the plasma with the Mg film, forming MgF and reducing the adhesion of SiNx during plasma deposition. It is also possible that removing the sample from the PECVD chamber multiple times could result in a film with higher oxynitride formation at the exposed faces and thus less compressive strain.

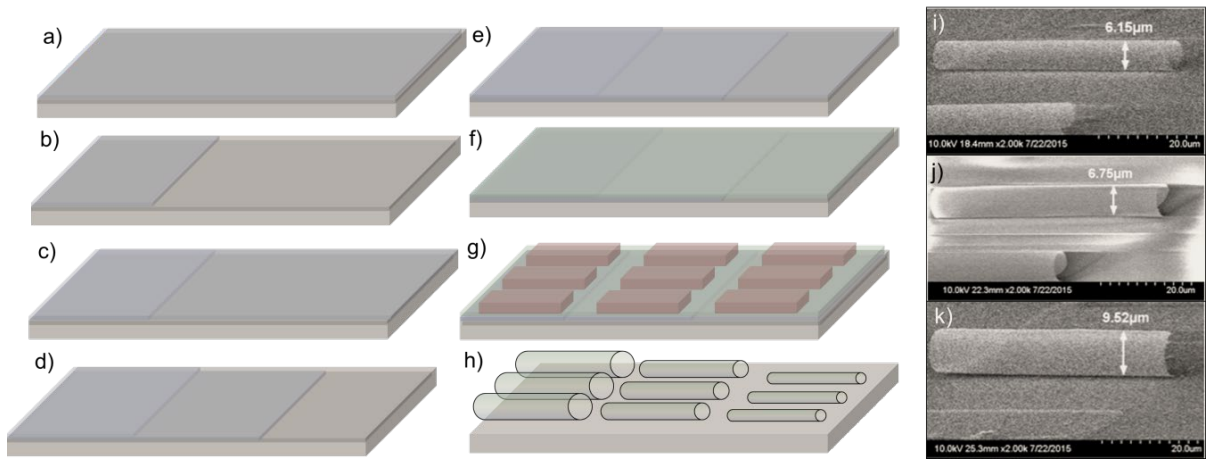

**Figure 2.** Multi-diameter micro-tubes on single substrate. The fabrication scheme for multi-diameter chips can be seen from (a)-(h): (a) deposition of the compressive layer, (b) patterned etch back of compressive layer, repetition of (a)-(b) until intended number different diameters is formed (c-e), (f) deposition of tensile layer, (g) mesa formation, and (h) sacrificial layer removal. The triple diameter chips are shown through SEM from thinnest bilayer (i) to thickest bilayer (k), and thus smallest and largest diameter micro-tubes, respectively.

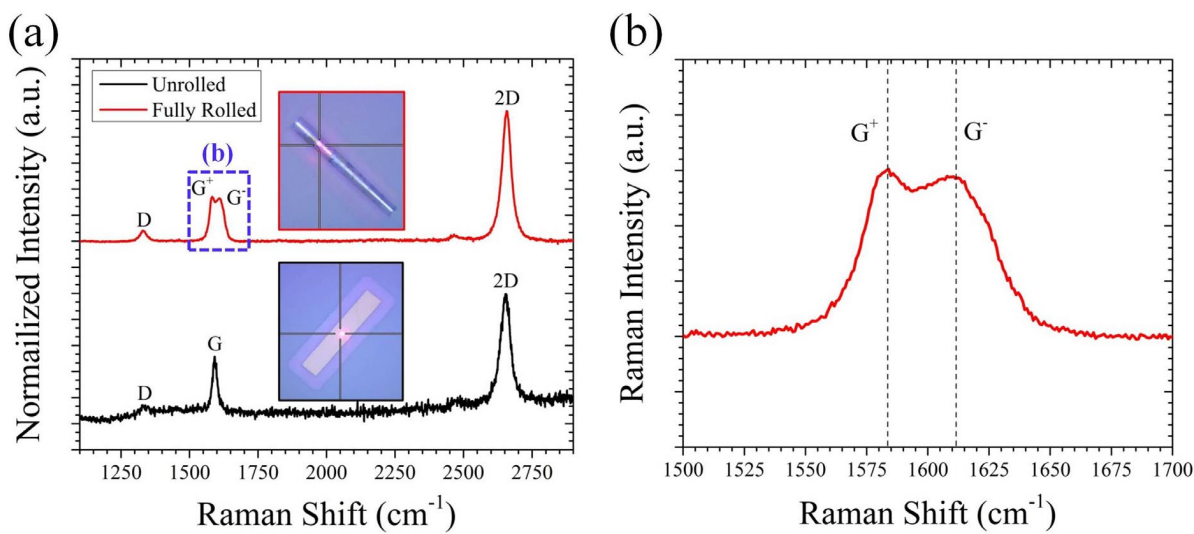

**Figure 3.** (a) Raman spectra obtained from flat (black curve) and fully rolled (red curve) graphene films. Optical images of the flat (black border) and fully rolled (red border) graphene/SiN$_x$ heterostructured micro-tubes are shown as insets, with the laser excitation spot location denoted by the intersecting cross-hairs. (b) Expanded view of doubly-degenerate G-band of the rolled graphene film, clearly showing the sub-band splitting (indicated by blue, dashed box in (a)).

Strain analysis was carried out after rolling of the graphene/SiN$_x$ micro-tube heterostructures via room-temperature micro-Raman spectroscopy, using a Renishaw inVia microscope system. Upon increasing compressive strain, the doubly degenerate G band of the graphene Raman spectrum becomes split into G$^+$ and G$^-$ sub-bands, the separation of which allows for quantifiable strain characterization [29]. Figure 3(a) shows representative Raman spectra obtained from flat (black curve) and fully rolled (red curve) graphene films. The graphene Raman signature shows a clear splitting of the G-band in the case of the rolled (compressively strained) foil, whereas no such sub-band signal is detected from the planar graphene sheet. An expanded view of the G-band signature of the strained graphene film is shown in Figure 3(b).

Interestingly, as rolling progresses by extending the duration of the sacrificial layer wet-etching period, the formation of narrower diameter tubes becomes associated with a characteristic increase in the separation of the G$^+$ and G$^-$ sub-bands of graphene Raman signature. Figure 4(a) plots the position of the G$^+$ and G$^-$ sub-bands as a function of the micro-tube diameter. Similarly, the micro-tube diameter-dependent G sub-band spacing is plotted in Figure 4(b), with the right-side vertical axis quantifying the resultant compressive strain in the graphene layer, as measured according to the formalism presented by Frank et al. [7].

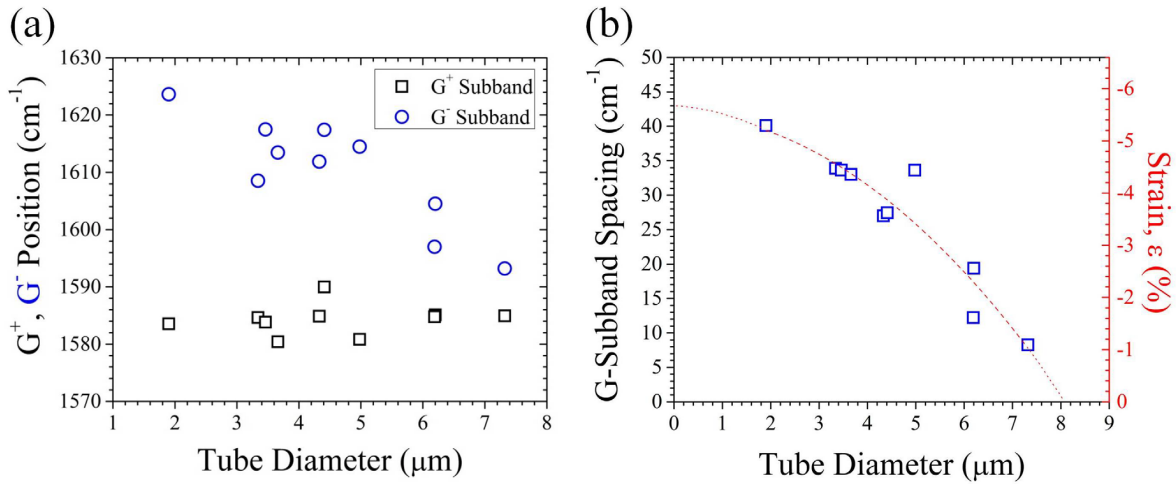

**Figure 4.** (a) Position of G$^+$ and G$^-$ sub-bands and (b) G sub-band spacing and graphene strain measured as function of the graphene/SiN$_x$ micro-tube diameter. The red curve in (b) shows a polynomial fit to the data, projected beyond the data set for smaller/larger tubes.

We note that the transferred graphene monolayer is attached to the stressed SiN$_x$ thin film stack by van der Waals force. During the processing steps from transfer to rolling, if the graphene layer is wrinkled or broken, it will not be conformal to the SiNx surface, hence not be in sync locally with the curvature of the rolled-up SiNx film stack. We indeed had samples that experienced processing issues and the trend observed in Figure 4 was not observed in those cases.

In summary, across a widely tunable micro-tube diameter range of roughly 1 – 8 microns, we have shown that over 5 % compressive strain can be induced in a mono-layer graphene film via spontaneous rolling in a $SiN_x$ bilayer structure. This work shows an unprecedented range of strain engineering in graphene and holds promise for similar strain control in other two-dimensional materials, such as $MoS_2$, wherein uniaxial deformation can allow for bandgap energy tunability. The self-rolled, three-dimension heterostructure platform highlighted here has potential for applications in ultra-compact inductors, micro-fluidic channel bio-sensors, rolled-up transistors, and micro-cavity optical resonators.